\documentclass[aps,showpacs,prd,twocolumn]{revtex4}
\usepackage{graphicx}
\usepackage{amssymb}
\usepackage{amsmath}
\newcommand{\be}{\begin{equation}}
\newcommand{\ee}{\end{equation}}
\newcommand{\ben}{\begin{eqnarray}}
\newcommand{\een}{\end{eqnarray}}
\newcommand{\bes}{\begin{subequations}}
\newcommand{\ees}{\end{subequations}}
\newcommand{\bb}{\bibitem}

\begin{document}

\title{Supersymmetry, shape invariance and the Legendre equations}
\author{D. Bazeia$^1$ and Ashok Das$^{2,3}$}
\affiliation{$^1$ Departamento de F\'{\i}sica, Universidade Federal da Para\'{\i}ba,
58051-900 Jo\~{a}o Pessoa, Para\'{\i}ba, Brazil.}
\affiliation{$^2$Department of Physics, University of Rochester, Rochester, New York 14627, USA}
\affiliation{$^3$Saha Institute of Nuclear Physics, 1/AF Bidhannagar, Calcutta 700064, India}


\begin{abstract}
In three space dimensions, when a physical system possesses spherical symmetry, the dynamical equations automatically lead to the Legendre and the associated Legendre equations, with the respective orthogonal polynomials as their standard solutions. This is a very general and important result  and appears in many problems in physics (for example, the multipole expansion etc). We study these equations from an operator point of view, much like the harmonic oscillator, and show that there is an underlying shape invariance symmetry in these systems responsible for their solubility. We bring out various interesting features resulting from this analysis from the shape invariance point of view. 
\end{abstract}

\maketitle

Supersymmetry has been shown to be an important tool for understanding physical systems of current interest.  It is mandatory for superstrings, and it finds applications in diverse areas of physics. A very nice example of this concerns the study of topological defects: Bogomol'nyi \cite{B} and Prasad and Sommerfield \cite{PS} have shown that under special circumstances, kinks, vortices and monopoles appear as stable, minimal energy states that solve first-order differential equations, forming BPS configurations in a supersymmetric environment. The BPS states are of current interest, since they offer different lines of investigations in physics. For instance, in the case of kinks in two-dimensional space-time, the study of classical stability of BPS states nicely leads us to supersymmetric quantum mechanics (SUSYQM) \cite{cooper}. There is also the inverse route  where we can re-construct BPS states starting with a given SUSYQM system, see, e.g., Refs.\cite{susy1,susy2,susy3}.

For conventional Hermitian Hamiltonians (of the type $H = p^{2} + V (x)$ with $2m=1=\hbar$ in one dimension), the SUSYQM problem is described by the first-order operators
\be
A = \frac{d}{dx}+W(x),\quad A^{\dagger} = - \frac{d}{dx}+W(x).\label{1}
\ee
Here the real function $W (x)$ is known as the superpotential, and with the above operators, we can define two  Hermitian Hamiltonians
\begin{align}
H_{-} & =A^{\dagger} A = -\frac{d^2}{dx^2}+ V_{-}(x),\nonumber\\
H_{+} & =A A^{\dagger} = -\frac{d^2}{dx^2}+ V_{+}(x),\nonumber\\
V_{\mp}(x) & =W^2 (x)\mp\frac{dW (x)}{dx}.\label{2}
\end{align}
These two Hamiltonians are supersymmetric partner Hamiltonians and share all the eigenvalues except for the ground state of $H_{-}$ which is assumed to have a vanishing eigenvalue (almost isospectral). The lowering and raising operators, $A, A^{\dagger}$, also take eigenstates of $H_{-}$ to those of $H_{+}$ and {\em vice versa}. Such a construction can also be carried out for non-Hermitian Hamiltonians \cite{nc}.

\begin{figure}[ht!]
\begin{center}
\includegraphics[scale=1]{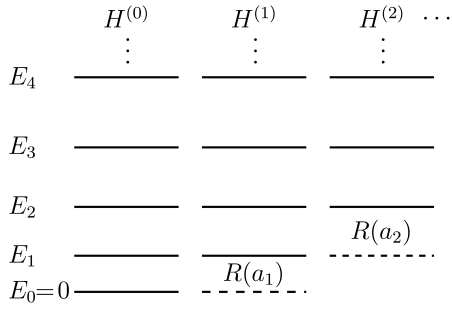}
\end{center}
\caption{Spectrum for a shape invariant family of Hamiltonians. The dashed lines represent the states that the partner Hamiltonians do not share.}
\label{f1}
\end{figure}

Supersymmetry relates only a pair of Hamiltonians and says that they are almost isospectral without actually determining the spectrum. On the other hand, shape invariance \cite{S1,S2} leads to a sequence of Hamiltonians which are pairwise supersymmetric. If the potential depends on a real parameter (e.g. a coupling constant) in such a way that,
\be
V_+(x,a_0) = V_- (x,a_1)+ R(a_1),\label{3}
\ee
where $a_{1} = f(a_{0})$ is a given function of the parameter $a_{0}$ and $R (a_{1})$, representing the spacing between the ground states of the two partner Hamiltonians, is a known (constant) function of the parameter, the potentials (Hamiltonians) are said to be shape invariant. In this case, we can construct a sequence of Hamiltonians \cite{ft}
\begin{align}
H^{(0)} & = H_{-} (a_{0}),\nonumber\\ 
H^{(1)} & = H_{+} (a_{0}) = H_{-} (a_{1}) + R (a_{1}),\nonumber\\
H^{(2)} & = H_{-}(a_{2})+ R(a_{1}) + R (a_{2}), \cdots\cdots ,\nonumber\\ 
H^{(k)} & = H_{-}(a_{k}) + \sum\limits_{j=1}^{k} R (a_{j}), \cdots,\label{4}
\end{align} 
all of which will be pairwise supersymmetric as shown in Fig. \ref{f1}. Shape invariance is a powerful symmetry which leads to a determination of the spectrum of the system. For example, it follows from \eqref{4} that the  $n$ th energy level of the original Hamiltonian $H_{-} (a_{0})$ is given by (the sum of the differences in the energy levels)
\be
E_{n} = \sum\limits_{k=1}^{n} R (a_{k}).\label{5}
\ee

Shape invariance has been exploited quite successfully in various areas of physics (see, e.g., Refs.~\cite{S3,S4,S5,S6,S7,S8,S9}). Harmonic oscillator is the simplest example of a shape invariant system. In this case, the parameter $a_{0} = \omega$ which does not change in the sequence of Hamiltonians (namely, $a_{k} = f (a_{k-1}) = a_{k-1} = \omega, k=1,2,\cdots$), only the potential (Hamiltonian) shifts leading to (the shift denoting the spacing between energy levels is a  constant independent of the level, $R (a_{k}) = a_{k} = \omega$)
\begin{align}
H^{(k)}  & = H^{(0)} + \sum\limits_{j=1}^{k} \omega = H^{(k-1)} + \omega,\nonumber\\ 
E_{n} & = \sum\limits_{k=1}^{n} \omega = \omega n.\label{6}
\end{align}
As one knows, the harmonic oscillator is described by the Hermite equation and this analysis shows that shape invariance is the underlying symmetry responsible for its solubility. In a similar spirit, we can ask if equations describing  other orthogonal polynomials may also have an underlying shape invariance symmetry responsible for their nice properties. In this letter we focus only on the Legendre and the associated Legendre equations and analyze the shape invariance associated with such systems. As we will see, the operators associated with the Legendre equation are not of conventional form and, therefore, the discussion of supersymmetry and shape invariance generalizes slightly. Since these equations as well as their regular solutions, namely, the Legendre and the associated Legendre polynomials appear in many areas of physics whenever we have a system with spherical symmetry (for example, in the multipole expansions etc), this analysis is of general and current interest.

The  Legendre equation can be written as ($x=\cos\theta, -1\leq x\leq 1$)
\be
\left(\frac{d}{dx} (x^{2} -1)\frac{d}{dx} - n (n+1)\right)P_{n} (x) = 0,\label{7}
\ee
where $n=0,1,2,\cdots$. For simplicity, we define 

\be 
H = \frac{d}{dx} (x^{2} -1)\frac{d}{dx},\label{7a}
\ee
so that we can write \eqref{7} also as
\be
\left(H - n(n+1)\right)P_{n} = 0,\label{8}
\ee 
From the recursion relations satisfied by the Legendre polynomials, we can determine the raising and lowering operators for the Legendre polynomials to be
\begin{align}
a_{n}^{\dagger} & = (x^{2}-1) \frac{d}{dx} + nx = (x^{2}-1)^{-\frac{n}{2} + 1} \frac{d}{dx} (x^{2}-1)^{\frac{n}{2}},\nonumber\\
a_{n} & = - \frac{d}{dx} (x^{2}-1) + nx = - (x^{2}-1)^{\frac{n}{2}} \frac{d}{dx} (x^{2}-1)^{-\frac{n}{2}+1},\label{9}
\end{align}
so that we have (for $n\geq 1$ and $P_{0}=1$)
\be
P_{n} = \frac{a_{n}^{\dagger}}{n}\, P_{n-1},\quad P_{n-1} = \frac{a_{n+2}}{n}\, P_{n},\label{10}
\ee
and satisfy the equation
\be
(a_{n+2}a_{n}^{\dagger} - n^{2})P_{n-1} = 0 = (1- x^{2})\left(H - (n-1)n\right)P_{n-1}.\label{11}
\ee
This equation can also be written equivalently as $(a_{n-1}^{\dagger}a_{n+1} - (n-1)^{2})P_{n-1} =0$ and would generate the polynomials $P_{-n-1}$ which are not independent so that we will continue with \eqref{11}. In other words, these polynomials indeed satisfy the Legendre equation. We note here that, unlike the harmonic oscillator system, here the forms of the raising and lowering operators depend on the level leading to a nonuniform ($n$ dependent) spacing between the levels as we will see.

The Legendre polynomials can be iteratively constructed from \eqref{10} and with the help of \eqref{9} we obtain a closed form expression of the form ($n\geq 1$)
\be
P_{n} = \frac{1}{n!}\,(x^{2}-1)^{-\frac{n}{2} + 1} \frac{d}{dx} \left((x^{2}-1)^{\frac{3}{2}} \frac{d}{dx}\right)^{n-1} (x^{2}-1)^{\frac{1}{2}},\label{12}
\ee
where we have used $P_{0} =1$. This form of the polynomial directly leads to the recursion relations, but seems quite different from the Rodrigues' formula \cite{rodrigues} which also gives a closed form expression for these polynomials. For low values of $n$, one can directly check that  the two expressions are equivalent. To show equivalence in general, let us assume that the two expressions are the same up to $P_{n-1}$ so that we can write \eqref{12}, for $n\rightarrow n-1$,  also as
\be
P_{n-1} = \frac{1}{2^{n-1} ((n-1)!)} \frac{d^{n-1} (x^{2}-1)^{n-1}}{dx^{n-1}}.\label{13}
\ee 
Then, using the identity
\be
(x^{2}-1) \frac{d^{n} (x^{2}-1)^{n-1}}{dx^{n}} = (n-1) n\,\frac{d^{n-2} (x^{2}-1)^{n-1}}{dx^{n-2}},\label{14}
\ee
it can be shown that \eqref{10} leads to
\begin{align}
P_{n} & = \frac{1}{2^{n-1} (n!)} (x^{2}-1)^{-\frac{n}{2}+1} \frac{d}{dx} (x^{2}-1)^{\frac{n}{2}}\frac{d^{n-1} (x^{2}-1)^{n-1}}{dx^{n-1}}\nonumber\\
& = \frac{1}{2^{n} (n!)} \frac{d^{n} (x^{2}-1)^{n}}{dx^{n}},\label{15}
\end{align}
so that the expression \eqref{12} is indeed equivalent to the Rodrigues' formula, but may be more useful in some cases.

The supersymmetry of the system is easily seen by noting from \eqref{11} that if
\be
a_{n+2} a_{n}^{\dagger} |\psi\rangle = E |\psi\rangle,\label{16}
\ee
then for $E\neq 0$ (or $n\neq 0$),
\be
a_{n}^{\dagger} a_{n+2} \left(a_{n}^{\dagger} |\psi\rangle\right) = E \left(a_{n}^{\dagger} |\psi\rangle\right),\label{17}
\ee
so that this pair of Hamiltonians can be thought of as supersymmetric partner Hamiltonians. However, the disadvantage of this description is that these Hamiltonians are not Hermitian. A Hermitian description is not necessary, but is much more desirable from the point of view of supersymmetry and shape invariance and this can be done in the following way.

Let us note from \eqref{11} that we can define a sequence of Hermitian Hamiltonians of the form 
\be
H^{(k-1)} = (1-x^{2})^{-1} \left(a_{k+2}a_{k}^{\dagger} - k^{2}\right) = H - (k-1)k,\label{18}
\ee
where $k=1,2,\cdots$ so that $H^{(0)} = H = \frac{d}{dx} (x^{2}-1) \frac{d}{dx}$. Since $H$ does not contain any potential term, the Hamiltonians in \eqref{18} are shape invariant with only a shift, as in the case of the oscillator, but in this case the shift depends on the level $k$. We note here that although the left hand side of \eqref{18} does not appear to be manifestly Hermitian, it really is, following from the identities
\begin{align}
(1-x^{2})^{-m} a_{n} & = a_{n-2m} (1-x^{2})^{-m},\nonumber\\ 
(1-x^{2})^{-m} a_{n}^{\dagger} & = a_{n+2m}^{\dagger} (1-x^{2})^{-m},\label{19}
\end{align}
where $|m|=0,\frac{1}{2}, 1, \cdots , \frac{n}{2}$. These identities are indeed quite useful in an algebraic study of this system. In fact, using \eqref{19}, we can also write $H^{(k-1)}$ in the manifestly Hermitian form
\be
H^{(k-1)} = (1-x^{2})^{-\frac{1}{2}}\left(a_{k+1}a_{k+1}^{\dagger} - k^{2}\right)(1-x^{2})^{-\frac{1}{2}}.\label{20}
\ee
The sequence of shape invariant Hamiltonians is now easily seen to satisfy 
\begin{align}
& (1-x^{2})^{-\frac{1}{2}} \left(a_{k+1}^{\dagger} a_{k+1} - k^{2}\right)(1-x^{2})^{-\frac{1}{2}}\nonumber\\
&\quad = (1-x^{2})^{-\frac{1}{2}} \left(a_{k+2} a_{k+2}^{\dagger} - (k+1)^{2}\right)(1-x^{2})^{-\frac{1}{2}}\nonumber\\
&\quad = H^{(k)} = H^{(k-1)} - 2 k.\label{21}
\end{align}
As we have mentioned before, here the order $k$ in the sequence (level) becomes the relevant parameter and comparing with the harmonic oscillator, we note that the shift between two levels now depends on the level. This has its origin in the $k$-dependent commutation relation
\be
[a_{k+1}, a_{k+1}^{\dagger}] = 2 k (1-x^{2}),\label{22}
\ee
which can be contrasted with the universal (level independent) commutator
\be
[a, a^{\dagger}] = 1,\label{23}
\ee
for the harmonic oscillator. The $n$ th level of $H = H^{(0)}$ can now be determined to be
\be
E_{n} = 2 \sum\limits_{k=1}^{n} k = n (n+1),\label{24}
\ee
as expected. This analysis brings out the intimate relationship between shape invariance and the solubility of the Legendre equation.

Let us next consider the associated Legendre equation which can be written as
\be
\left(H + \frac{m^{2}}{1-x^{2}}\right) P_{n,m} (x) = n (n+1) P_{n,m} (x),\label{25}
\ee
where $m = 0, \pm 1, \pm 2,\cdots , \pm n$ and $H$ is defined in \eqref{7a}. This equation is symmetric under $m \leftrightarrow - m$ and so we concentrate only on the values $m\geq 0$. We note that for $m=0$, this equation reduces to the Legendre equation in \eqref{8}. To analyze the supersymmetry and shape invariance associated with the associated Legendre equation, let us define the raising and lowering operators
\begin{align}
A_{m}^{\dagger} & = (1-x^{2})^{\frac{m+1}{2}} \frac{d}{dx} (1 - x^{2})^{-\frac{m}{2}}\nonumber\\
& = (1-x^{2})^{\frac{1}{2}} \frac{d}{dx} + \frac{mx}{(1-x^{2})^{\frac{1}{2}}},\nonumber\\
A_{m} & = - (1-x^{2})^{- \frac{m}{2}} \frac{d}{dx} (1 - x^{2})^{\frac{m+1}{2}}\nonumber\\
& = - \frac{d}{dx}(1-x^{2})^{\frac{1}{2}} + \frac{mx}{(1-x^{2})^{\frac{1}{2}}},\label{26}
\end{align}
so that we have the recursion relation
\be
P_{n,m+1} = A_{m}^{\dagger} P_{n,m}.\label{26a}
\ee
With these we can now write (see \eqref{25})
\be
H^{(m)} = H + \frac{m^{2}}{1-x^{2}} = A_{m}A_{m}^{\dagger} + m (m+1),\label{27}
\ee
where $H^{(0)} = H$ is defined in \eqref{7a}. We note that for any $E\neq 0$, if 
\be
\left(A_{m} A_{m}^{\dagger} + m (m+1)\right) |\psi\rangle = E |\psi\rangle,\label{28}
\ee
then,
\be
\left(A_{m}^{\dagger} A_{m} + m(m+1)\right) \left(A_{m}^{\dagger} |\psi\rangle\right) = E \left(A_{m}^{\dagger} |\psi\rangle\right),\label{29}
\ee
so that these can be thought of as supersymmetric partner Hamiltonians. In fact, it is straightforward to show that (see \eqref{27})
\begin{align}
A_{m}^{\dagger} A_{m} & + m (m+1)  = A_{m+1} A_{m+1}^{\dagger} + (m+1)(m+2)\nonumber\\
& = H^{(m+1)} = H + \frac{(m+1)^{2}}{1-x^{2}}.\label{30}
\end{align}

This shows that at any level of $H$, namely, $E_{n} = n (n+1), n\neq 0$, we can construct a sequence of shape invariant Hamiltonians given by
\be
H^{(m)} = H + \frac{m^{2}}{1-x^{2}},\quad m = 0, 1, 2,\cdots, n,\label{31}
\ee
where the parameter on which the potential depends can be identified with $a_{m} = m, a_{m+1} = f (a_{m}) = m+1$. Furthermore, this is an example of shape invariance without any shift in the Hamiltonians, $R(a_{m}) =0$, so that all the $(2n+1)$ states (including the states with $m<0$) will have degenerate eigenvalues. Degeneracy in the azimuthal quantum numbers is normally understood as a consequence of rotational symmetry in a system. However, here we have a new interpretation of this degeneracy from the point of view of shape invariance of the system.

An immediate consequence of this shape invariance analysis is that we can relate the associated Legendre polynomials, $P_{n,m}$, directly to the Legendre polynomials through the lowering operators. Constructing the wave functions iteratively using \eqref{26a} we obtain 
\begin{align}
P_{n,m} (x) & = A_{m-1}^{\dagger} A_{m-2}^{\dagger}\cdots A_{1}^{\dagger}A_{0}^{\dagger} P_{n} (x)\nonumber\\
& = (1-x^{2})^{\frac{m}{2}}\,\frac{d^{m}P_{n} (x)}{dx^{m}}\nonumber\\
& = \frac{1}{2^{n} (n!)}\, (1-x^{2})^{\frac{m}{2}} \frac{d^{n+m} (x^{2}-1)^{n}}{dx^{n+m}},\label{32}
\end{align}
where we have used the definitions in \eqref{26} as well as the Rodrigues' formula \eqref{15}.  These relations are well known in the literature, but shape invariance leads to a simple and direct derivation for them. We note here that sometimes the associated Legendre polynomials (for $m>0$) are defined with a Condon-Shortley phase $(-1)^{m}$ which can be easily incorporated into our formalism by changing the signs of the lowering and the raising operators in \eqref{26}. For completeness we note that the associated Legendre polynomials for $m <0$ are proportional to those with $m > 0$ and the exact relation is given by
\be
P_{n, -m} (x) = (-1)^{m}\,\frac{(n-m)!}{(n+m)!}\, P_{n,m} (x),\label{33}
\ee
which also follows from \eqref{32}.

To conclude we recall that the Legendre and the associated Legendre equations are fundamental in the study of physical systems with spherical symmetry and their solutions, the Legendre and the associated Legendre polynomials, have deep relations in physics. These equations have been studied exhaustively in the past. In this letter we have described a new way of looking at these equations and their solutions from the point of view of an underlying supersymmetry and shape invariance. This can be thought of as an operator analysis of the system, much like in the case of the harmonic oscillator, which is obtained by exploiting the underlying supersymmetry and shape invariance of the system. This analysis leads directly to a closed form expression for the Legendre polynomials which is equivalent to but is manifestly different from the Rodrigues' formula and which may be useful in some studies. Degeneracy of the azimuthal quantum numbers in the case of rotationally invariant systems finds a new interpretation from the point of view of shape invariance. Other properties such as the relation between the associated Legendre polynomials and the Legendre polynomials follow directly from such an operator analysis. There are, of course, other orthogonal polynomials which are also of importance in physics. It would be interesting to examine which of these are described by equations with an underlying shape invariance symmetry.

The authors would like to thank CNPq/Brazil for partial financial support.


\end{document}